\documentclass[Afour,sagev,times]{sagej}

\usepackage{moreverb,url}

\usepackage[colorlinks,bookmarksopen,bookmarksnumbered,citecolor=red,urlcolor=red]{hyperref}

\newcommand\BibTeX{{\rmfamily B\kern-.05em \textsc{i\kern-.025em b}\kern-.08em
T\kern-.1667em\lower.7ex\hbox{E}\kern-.125emX}}

\def\volumeyear{2016}

\begin{document}

\runninghead{Mendes-Neves et al.}

\title{The Geostrategy of Youth Player Recruitment in Portuguese Clubs}

\author{Tiago Mendes-Neves\affilnum{1,3}, Luís Meireles\affilnum{3}, João Mendes-Moreira\affilnum{1,2} and Nuno de Almeida\affilnum{4}}

\affiliation{\affilnum{1}Faculdade de Engenharia, Universidade do Porto, Porto, Portugal\\
\affilnum{2}LIAAD, INESC TEC, Porto, Portugal\\
\affilnum{3}Nordensa Football, Cluj, Romania\\
\affilnum{4}Independent Researcher, Porto, Portugal}

\corrauth{Tiago Mendes-Neves, Faculdade de Engenharia, Universidade do Porto, Porto, Portugal.}

\email{tiago.neves@fe.up.pt}

\begin{abstract}

Portugal's prominent role as a global exporter of football talent is primarily driven by youth academies. Notably, Portugal leads the global ranking in terms of net transfer balance. This study aims to uncover and understand the recruitment strategies of Portuguese clubs for sourcing young talent and evaluate the relative success of different strategies. A comprehensive dataset spanning recent decades of Portuguese youth and professional football provides granular insights, including information such as players' birthplaces and the initial grassroots clubs where they developed. The initial findings suggest a correlation between a club's prominence and the geographic reach of its youth scouting operations, with larger clubs able to cast their net wider. Analysis of the correlation between players' birthplace and high-tier football club location suggests that the performance of senior teams acts as a catalyst for investment in youth teams. Regions without professional clubs are often left underserved. That said, certain clubs have made significant gains by focusing on player recruitment outside their district, such as the Algarve region, demonstrating how geographically targeted strategies can deliver substantial returns on investment. This study underscores data's role in sharpening youth player recruitment operations at football clubs. Clubs have access to in-depth and comprehensive datasets that can be used for resource allocation, territorial coverage planning, and identifying strategic partnerships with other clubs, potentially influencing their future success both on the field and financially. This offers opportunities for growth for individual clubs and holds implications for the continued strength of Portuguese football.

\end{abstract}

\keywords{spatio-temporal analysis, youth intake, player recruitment, geostrategy, big data in soccer}

\maketitle

\section{Introduction}

Since the beginning of the century, the major clubs in Portugal, known as the Big Three (FC Porto, SL Benfica, and Sporting CP), have significantly invested in youth academies. While Sporting invested 17M€ in 2002 to build the now-denominated Academia Cristiano Ronaldo, Benfica invested 15M€ in 2005 to build Benfica Campus. While these investments include more than youth facilities, they are still substantial investments for the time we referred. Porto had the “611” project, an alternative approach that focused on improving the quality of human resources and standardizing how the academy worked across different levels.

On top of these investments, the annual maintenance cost of a youth academy like those referred to reaches up to 12M€\cite{benfica-academy-inside-joao-felix}. Such value includes staff salaries, maintenance, scouting, and players’ accommodation\cite{ford_survey_2020}.

\begin{figure}[h]
\centering
\includegraphics[width=\linewidth]{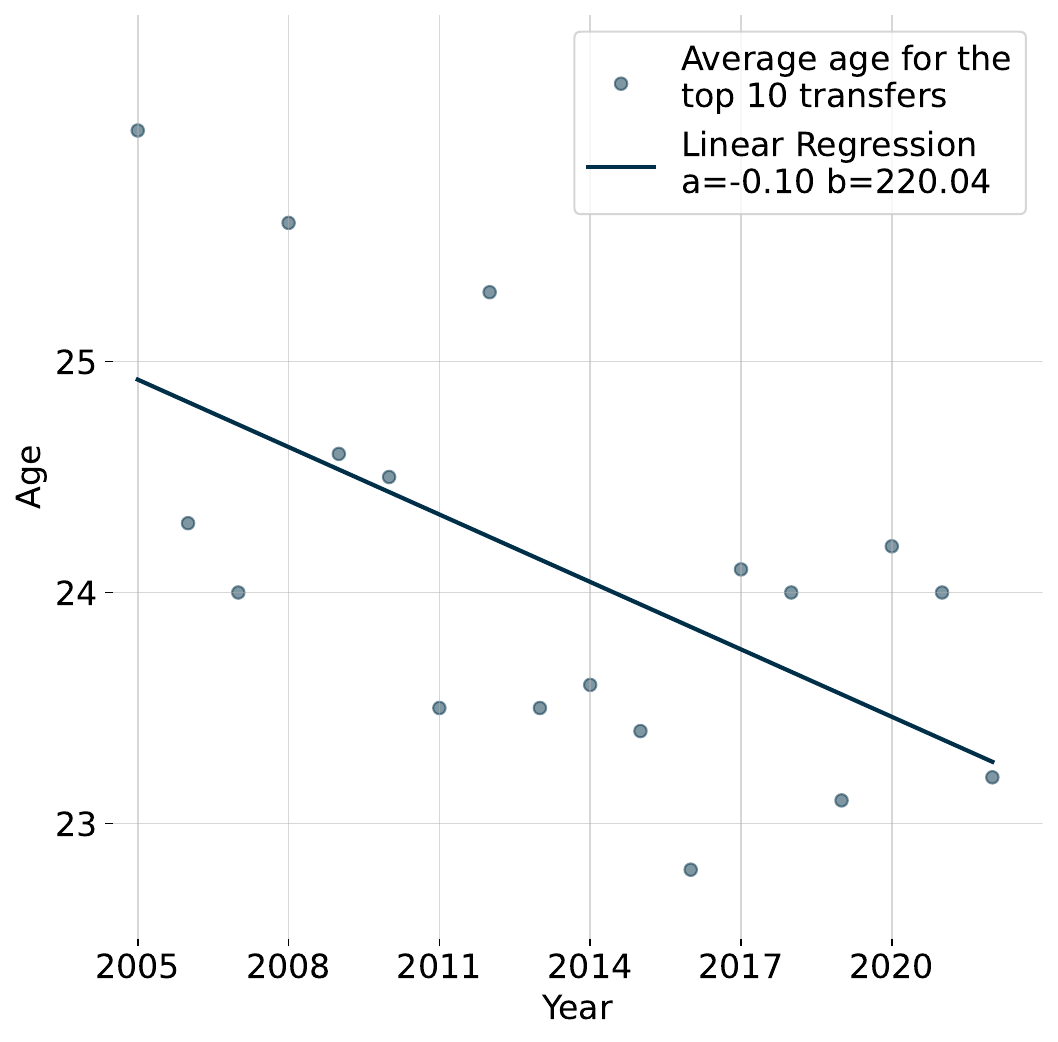}
\caption{The decrease in the age of the top 10 transferred players. Coefficients a and b are from a linear regression performed on the data displayed, taking the shape y=ax+b, where x is the year the player is transferred, and y is the player's age at the transfer time. The data used to produce this figure is described in the Methods section.}
\label{transfer_revenue_age}
\end{figure}

Nonetheless, these projects have been generating an incredible return on investment. For instance, in 2019, João Félix was transferred by 126M€ from SL Benfica to the Spanish side Club Atlético de Madrid. This single transfer provided revenue that exceeds all the investments Benfica made in its academy since 2005. And current trends can make these transfers look like a drop in the ocean through the next decade. Indeed, analyzing the last two decades’ summer transfer windows, we can see that European “big sharks” are willing to invest their money in ever younger football players (see Figure \ref{transfer_revenue_age}). 

\begin{figure}[h]
\centering
\includegraphics[width=\linewidth]{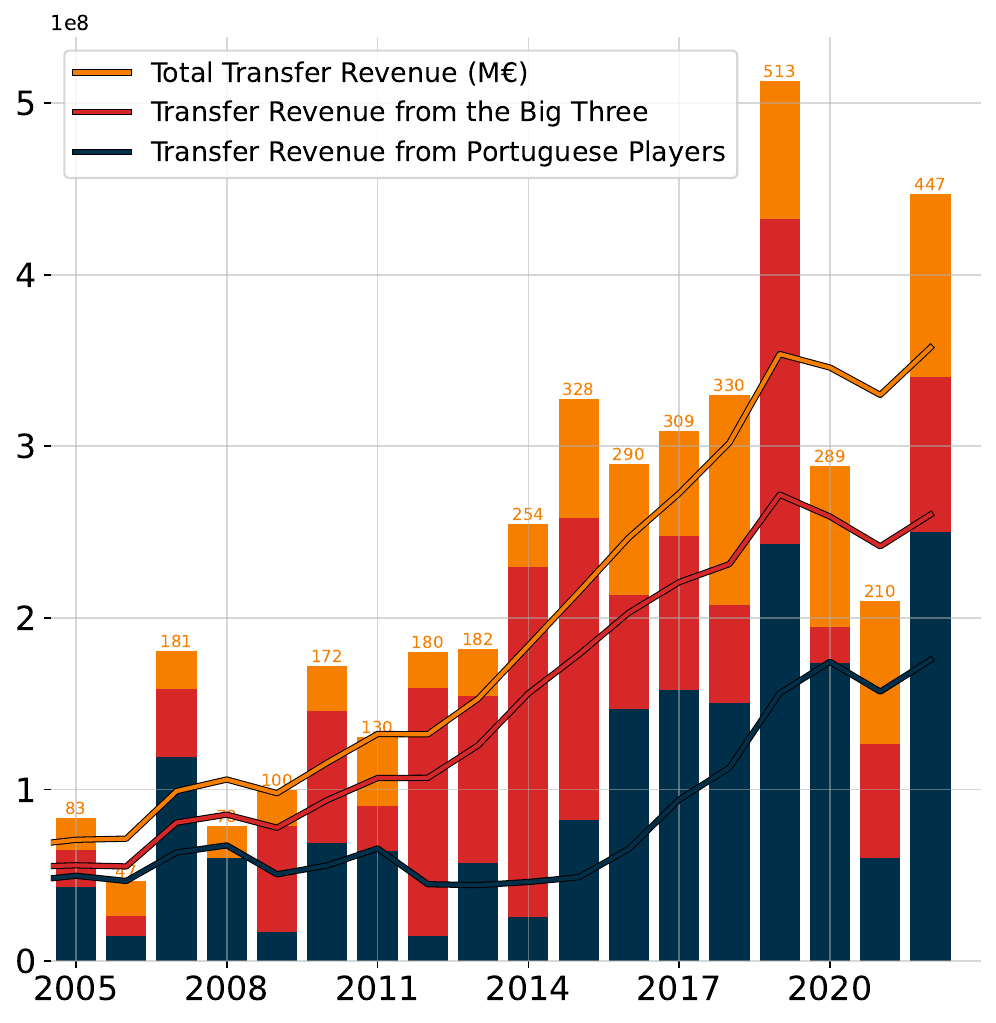}
\caption{The transfer revenue for the Portuguese Top Tier since 2005. Bars indicate the total amount received, and the lines are 5-year averages. Note that the bars displayed represent absolute values, not cumulative.}
\label{transfer_revenue_2}
\end{figure}

Many note that investment in football has skyrocketed in this century \cite{haugaasen_developing_2012,zulch_management_2020}. From television rights deals in the billions of dollars per season to private equity investments that turn modest teams into multi-billion dollar enterprises, income has risen substantially, allowing teams to invest heavily in the transfer market. As one man’s spending is another man’s income, transfer revenue increased at a \~12\% Compound Annual Growth Rate (CAGR) in the Portuguese first division. Even though the slump in the 2020/2021 and 2021/2022 seasons due to COVID hits to the sport, the revenue trendline is already back up to 2019/2020 levels, as visually presented in Figure \ref{transfer_revenue_2}.

\begin{figure}[h]
\centering
\includegraphics[width=\linewidth]{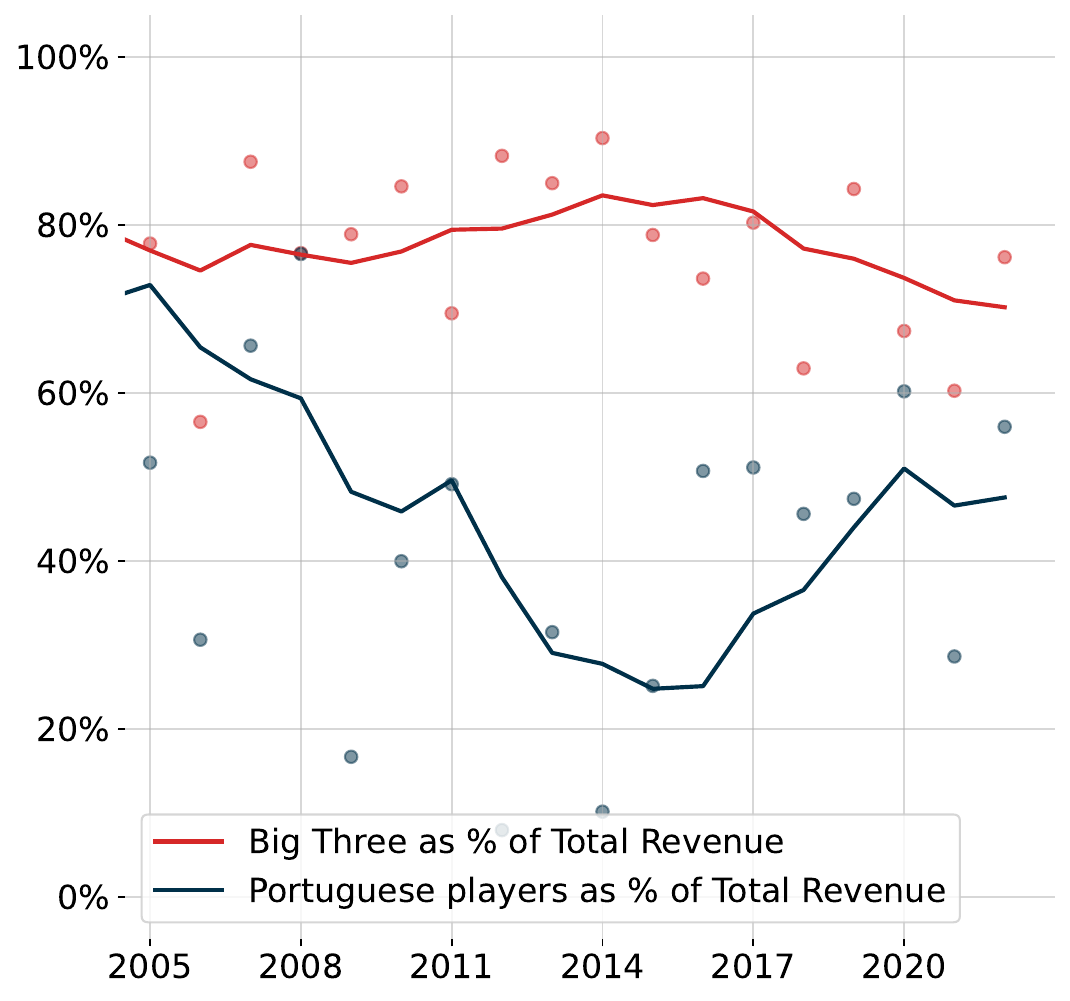}
\caption{The trends in the percentage of total revenue for the Big Three and Portuguese players.}
\label{transfer_revenue_3}
\end{figure}

Clubs are increasing their revenue substantially, and according to Figure \ref{transfer_revenue_3}, two trends are catching the eye: (1) clubs outside the Big Three and (2) Portuguese players are increasing their transfer revenue share. 

The first trend may be happening due to two factors: (1) on the one hand, the Big Three are not the only clubs to have invested in academies, and on the other, (2) the Big Three are not able to integrate all the players that pass through their youth academies in their professional teams, which are then appreciated at other clubs throughout the Portuguese pyramid. In the latter point, the Portuguese football pyramid benefits from the investment made by the biggest clubs in Portugal, providing them with talent that would otherwise have been developed less extensively.

The second trend indicates that investing in youth football provides excellent club returns. From the beginning of the century, many foreign players have made their way to the Portuguese first tier due to the effect of globalization on football and permissive legislation regarding the number of foreign players per squad. Such a trend is changing as investment in youth development starts providing returns to investors. As a result, Portugal is seen as a stepping stone for football players to jump to more competitive leagues or lucrative contracts.

A key component of youth academies is youth recruitment. In the race for talent, clubs are impelled to recruit young players as early as possible. The specialization process usually starts at the U15 level \cite{bloom_developing_1985, ericsson_role_1993}, i.e., players start focusing on learning the traits of a specific field role instead of understanding the general dynamics of the game. Clubs recruit players for their academies earlier, allowing them to monitor players and increase deliberate practice, one of the most important factors in talent development\cite{coutinho_talent_2016}. U12 scouts indicate that, on average, they can reliably predict a player's future performance at 13.6 years old\cite{bergkamp_how_2022}, with other studies showing that skill level at 15 years old are strong predictors of performance four years later\cite{forsman_identifying_2016}. Local academies are crucial to allow players to learn these skills at an early age, as players become harder to teach at an older age\cite{larkin_talent_2017}. Moreover, high-quality environments are linked to strong player development\cite{mills_examining_2014,pulido_sport_2018,kelly_multidisciplinary_2022}.

How clubs recruit for youth teams can tell us a lot about their overall strategy. Increasing the coverage of scouts leads to an improved player selection process, which benefits some perspectives. For example, from a genetic perspective\cite{mcauley_genetic_2021}, it increases the range of genetic profiles in the academy. From the psychological perspective, it allows the recruitment of players with diverse profiles, which plays a role in the transition to the first team\cite{finn_coping_2010,forsman_identifying_2016}.

For this paper, we created and analyzed territory maps containing information on where clubs recruit young players in Portugal, aiming to answer the following questions:

\begin{itemize}
    \item How are clubs recruiting, and what components of their strategy can we retrieve?
    \item What factors impact how much talent is generated in certain regions?
    \item What are some potentially underexplored regions in the country?
    \item How can smaller clubs take advantage of the investment (theirs and of their competitors) to compete against the Big Three?
\end{itemize}

The paper structure is as follows:

\begin{itemize}
    \item Section 2 presents the dataset used and the methods used to produce the results.
    \item Section 3 presents the results.
    \item Section 4 discusses the results.
    \item Section 5 presents the concluding remarks of the paper.
\end{itemize}

\section{Methods and Materials}
\subsection{Datasets}
This work comprises data obtained from several disparate sources:

\begin{itemize}
    \item Transfermarkt.pt: used to retrieve data regarding the transfer history of players. 
    \item Zerozero.pt: used to retrieve data concerning players’ birthplaces and registered competitive data from youth into senior competitions. We collected data for every team in the following leagues: Portuguese First-tier, Second-tier, Third-tier, Fourth-tier, the under-23 league, the first two tiers of the under-19 and under-17 levels, and the first tier of under-15.
    \item Wikipedia.pt: data regarding territorial density.
    \item FIFA Professional Football Report 2019 \cite{fifa_professional_football_department_professional_2019}.

\end{itemize}

This paper focuses exclusively on men's soccer. The data sources do not provide data from the women's soccer perspective, making the analysis impossible. It would be very important to invest in the organization of such data, as it is important to close the gap between both sports\cite{curran_what_2019}.

\subsection{Exploratory Data Analysis}
\begin{figure}[h]
\centering
\includegraphics[width=\linewidth]{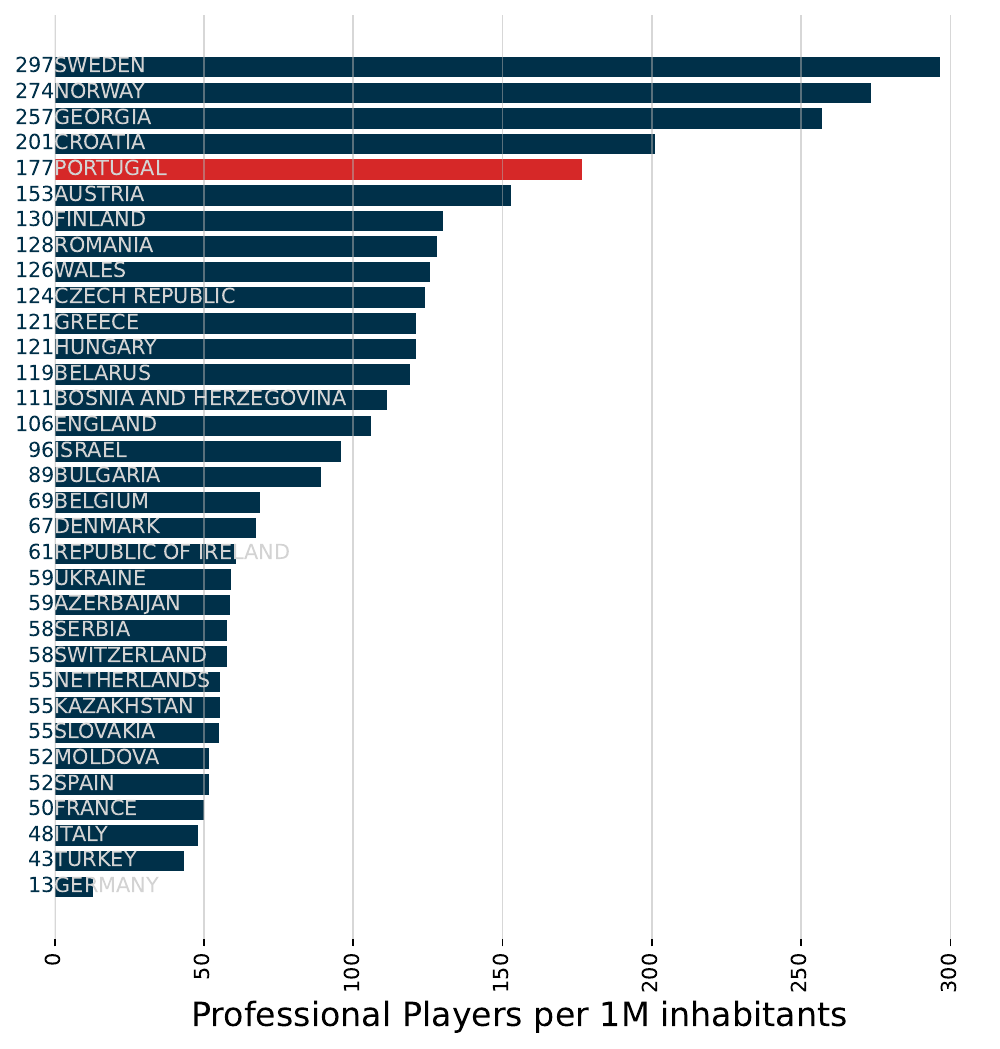}
\caption{Number of professional players per 1M inhabitants for all countries in UEFA with over 3M inhabitants \cite{fifa_professional_football_department_professional_2019}. Missing data from Russia, Poland, and Scotland.}
\label{players_per_cap}
\end{figure}

Before diving into the main questions, reflecting on why Portugal is a hotbed for talent is essential. The importance of the game is well represented in Figure \ref{players_per_cap}. Portugal ranks 5th as a European nation with a population of over 3M in terms of professional soccer players per capita. These numbers are only surpassed by two Nordic countries (Sweden and Norway), Georgia and Croatia. While the Nordic countries’ numbers are explained by the country's better economic environment, Portugal (and Georgia and Croatia) do not have the same conditions. The high amount of professional soccer players in Portugal is explained mainly by the cultural importance of the game \cite{nina_clara_tiesler_o_2006}.

\begin{figure*}[h]
\centering
\includegraphics[width=\linewidth]{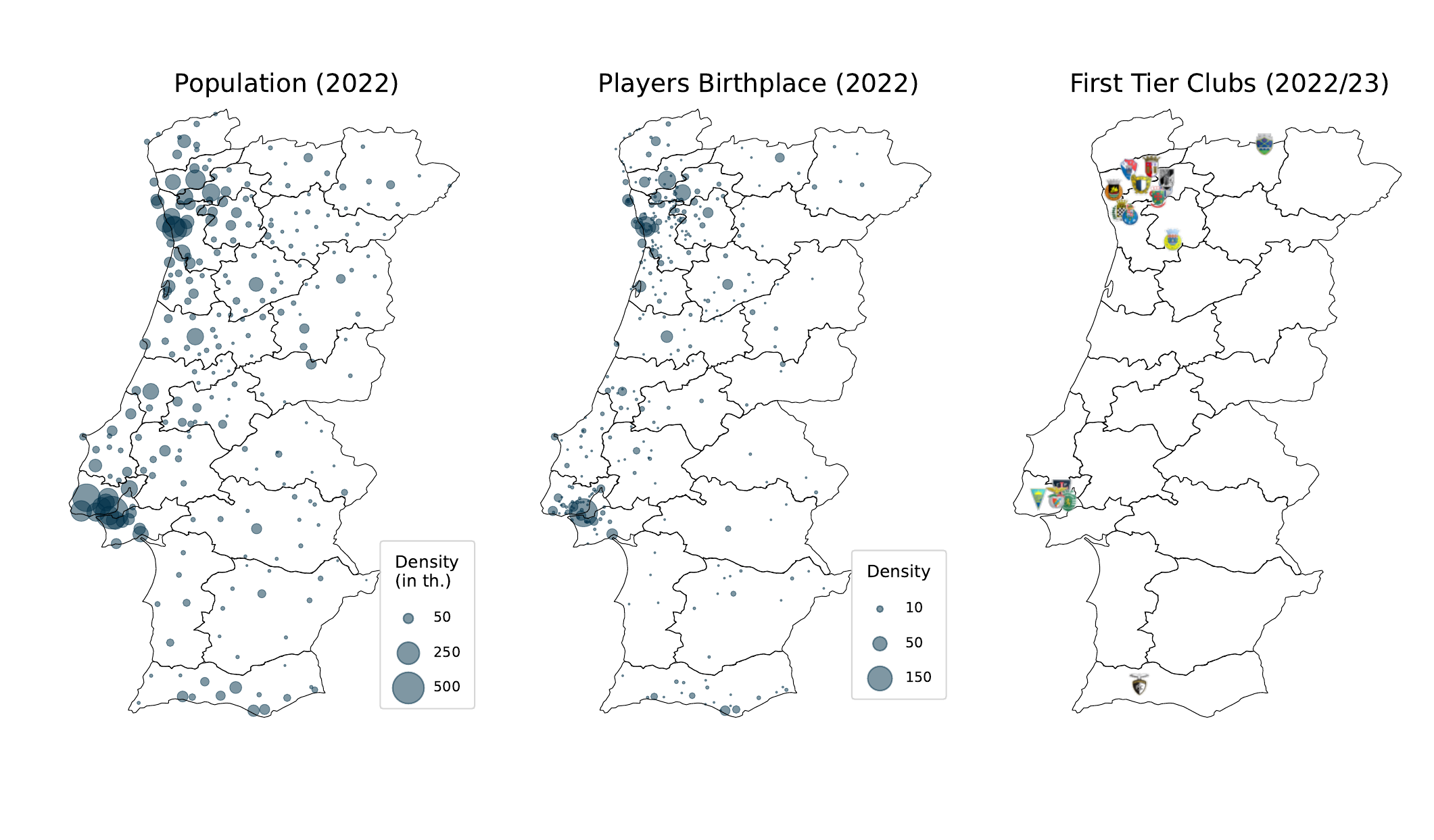}
\caption{The overall picture of Portugal’s population, professional player’s birthplace, and clubs in the top tier.}
\label{where}
\end{figure*}

Portugal also has some conditions that are rare in other countries, visible in Figure \ref{where}: (1) the population is highly centralized in two central regions (Lisbon and Northern Coastal), (2) most resources are centralized in the Big Three clubs, and (3) the teams in the top-tier are highly centralized towards the population centers. This combination of factors leads to efficiencies in running youth academies. Larger clubs recruit players from all over the country, and as they cannot absorb all the talent, it then flows toward surrounding smaller clubs.

\begin{figure}[h]
\centering
\includegraphics[width=\linewidth]{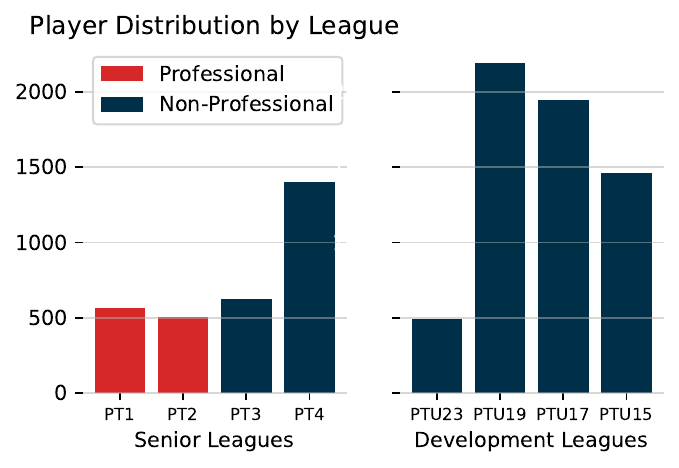}
\caption{The distribution of players across the different tiers in the Portuguese football pyramid, according to the data used in this work.}
\label{league_distribution}
\end{figure}

The football pyramid in Portugal comprises the top and second tiers as professional leagues. Nevertheless, most registered football players in Portugal are not professionals, playing either in lower divisions or youth academies. There are more players in youth academies than in senior leagues across the country, with the distribution presented in Figure \ref{league_distribution}.

\begin{figure}[h]
\centering
\includegraphics[width=\linewidth]{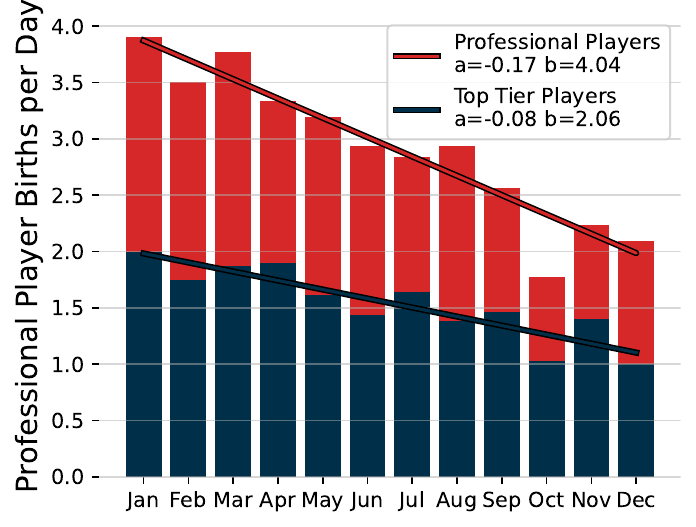}
\caption{The RAE for professional players in Portugal. Per month, professional player losses amount to 5.1 (0.17 times a 30-day average month), and top-tier players’ losses are 2.3. Coefficients a and b are from a linear regression y=ax+b, where x is the number of the month, and y is the number of professional players born per day.}
\label{birth_month_distribution}
\end{figure}

Several insights can be extracted from this data. For example, we can confirm the relative age effect (RAE) – i.e., the phenomenon of overrepresentation of players born in the first months of each year \cite{helsen_relative_2005} and even quantify it. Figure \ref{birth_month_distribution} presents the data on the professional leagues, and for the first two tiers combined, we lose on average 5.1 (0.48\% of the player base) soccer players per month, while the loss in the first tier is reduced to 2.3 (0.41\%). The data presented in Table 1 shows that the effect on lower levels is even more significant. This effect results in a loss of talent due to an inadequate selection process\cite{baker_compromising_2018,bennett_creating_2019}.

\begin{table}[h]
\small\sf\centering
\caption{The loss of players per month across the Portuguese football pyramid.}
\begin{tabular}{l|r|r|r}
\toprule
League & \# Samples & Change & Change (\%)\\
\midrule
PT1 & 562 & -2.32 & -0.41\% \\
PT2 & 504 & -2.81 & -0.56\% \\
PT3 & 625 & -3.25 & -0.52\% \\
PT4 & 1399 & -5.49 & -0.39\% \\
PTU15 & 1462 & -12.94 & -0.89\% \\
PTU17 & 808 & -8.23 & -1.02\% \\
PTU19 & 709 & -7.28 & -1.03\% \\
PTU23 & 488 & -3.69 & -0.76\% \\
PTU17 2nd & 1139 & -8.17 & -0.72\% \\
PTU19 2nd & 1481 & -6.17 & -0.42\% \\
\bottomrule
\end{tabular}
\end{table}

\begin{figure}[h]
\centering
\includegraphics[width=\linewidth]{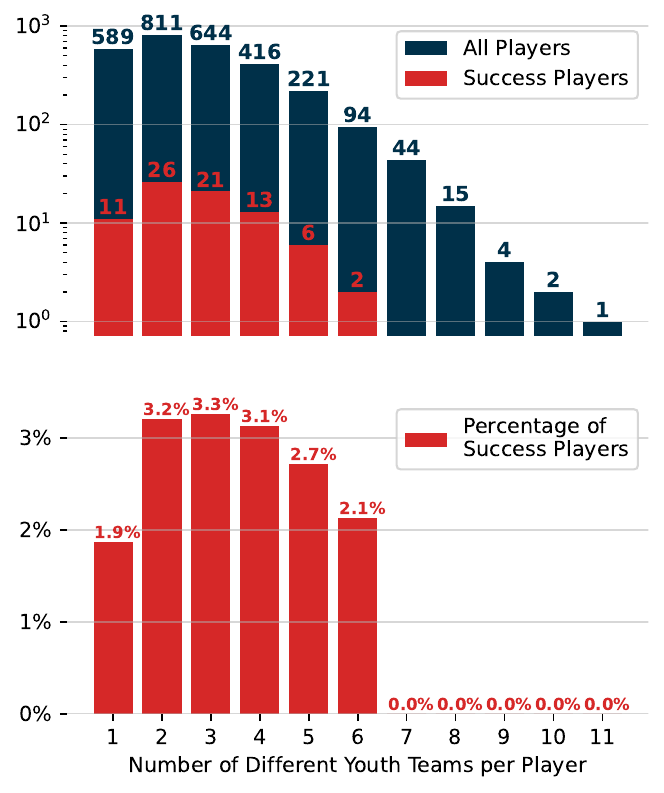}
\caption{The number of players that fulfill our success criterion. The bar chart on top indicates absolute values, while the bar chart on the bottom indicates the percentage of successful players related to total players.}
\label{youth_teams_distribution}
\end{figure}

In Portugal, a reasonable success measure for a football player is whether they played in one of the Big Three. Even though this criterion is stringent and limited since it misses international players, it provides a clean measure without requiring control for the national team’s quality. Bernardo Silva and Rafael Leão are good examples of players that do not fit this criterion, although they should. In Figure \ref{youth_teams_distribution}, we present how many players achieved this success criterion and the number of youth teams they played for in their formative years.

One of the key insights is that players benefit from changing teams during their youth years, but only up to a certain point. By being exposed to new challenges, players can improve their chances of making a successful career. The optimal point is three to four clubs during their youth years. We can also speculate, based on recent and related sports expertise development research, that the putative experience of some of these players being rejected or, for some other reason, having to move away from a specific academy where they were comfortable can work as a “traumatic trigger” for developing the needed psychological resilience and psychological robustness to strive as professionals \cite{van_yperen_why_2009,hardy_great_2017}.

Furthermore, we can also compare the youth academies’ performances, especially the most important - i.e., those from the Big Three. By measuring the impact of being enrolled in the youth academy of each Big Three for a given year, we can estimate how much playing in a specific academy improves the odds of a player meeting the successful player criterion. In essence, for every year, we calculate the linear regression coefficients between three binary input variables (played in Porto/Benfica/Sporting’s academy) and the binary output variable (is a successful player). 

\begin{figure}[h]
\centering
\includegraphics[width=\linewidth]{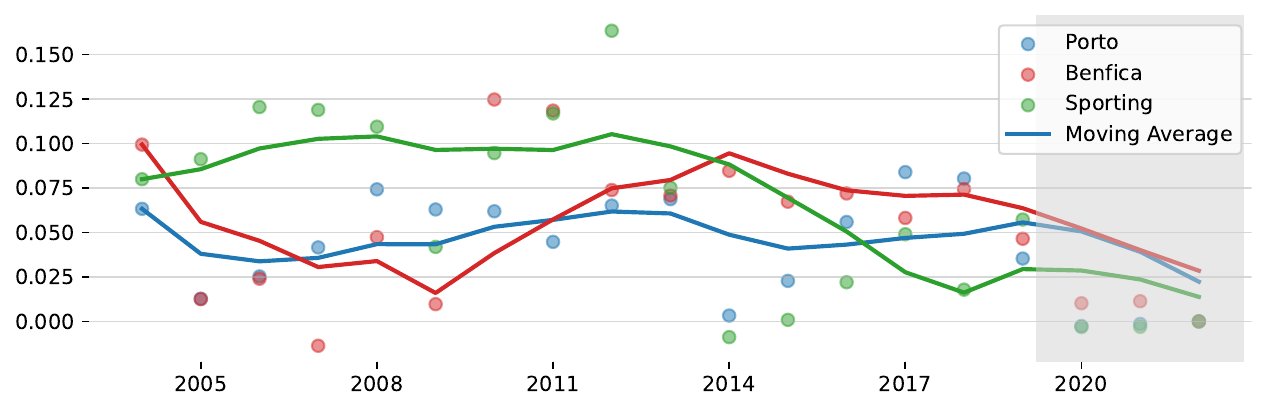}
\caption{The impact of youth academies on a player’s probability of success.}
\label{linear_regression_formacao}
\end{figure}

We present the results in Figure \ref{linear_regression_formacao}. Excluding the later years (i.e., 2015 onwards), since the current academy players are still on the journey to nurture as complete professionals, we can observe that Sporting CP took the lead during the 2004-2014 period regarding the abovementioned criteria. After that, we can observe a sharp decline in the amount of talent Sporting CP brought out of the academy into the first team. SL Benfica increased its ratings substantially during the early part of the 10s, surpassing Sporting in 2014 and leading the pack until the writing of this paper. It is interesting to note that FC Porto academy’s output has remained relatively consistent throughout the years, although approaching the second place in the later years.

\subsection{Club Intake Map}
\begin{figure}[h]
\centering
\includegraphics[width=\linewidth, trim={0 15cm 8cm 0}, clip]{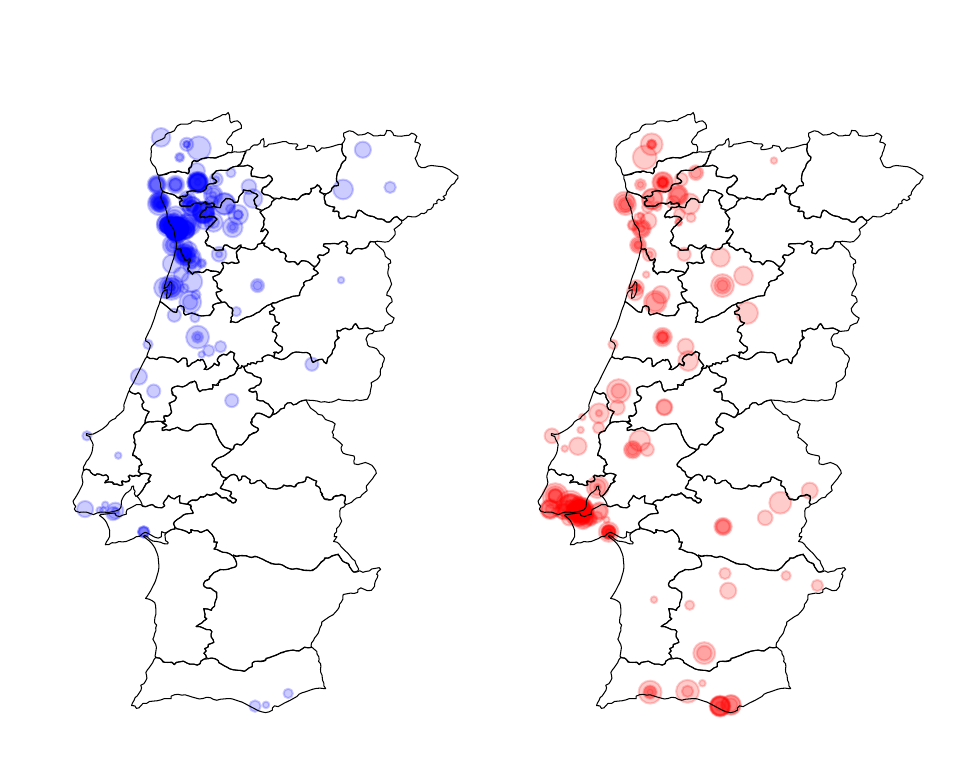}
\caption{A visualization of the inputs used to produce the KDE maps.}
\label{club_formacao}
\end{figure}

To build a club’s youth intake map, we used the following strategy:

\begin{itemize}
    \item For each season played in the club, add a data point to the location where the player was born. The resulting data is similar to the data presented in Figure \ref{club_formacao}.
    \item Using the collected data points, calculate the Kernel-Density Estimation (KDE) and plot it using Seaborn’s “kdeplot” function\cite{Waskom2021}.
\end{itemize}

\section{Results}

\begin{figure*}[h]
\centering
\includegraphics[width=0.8\linewidth]{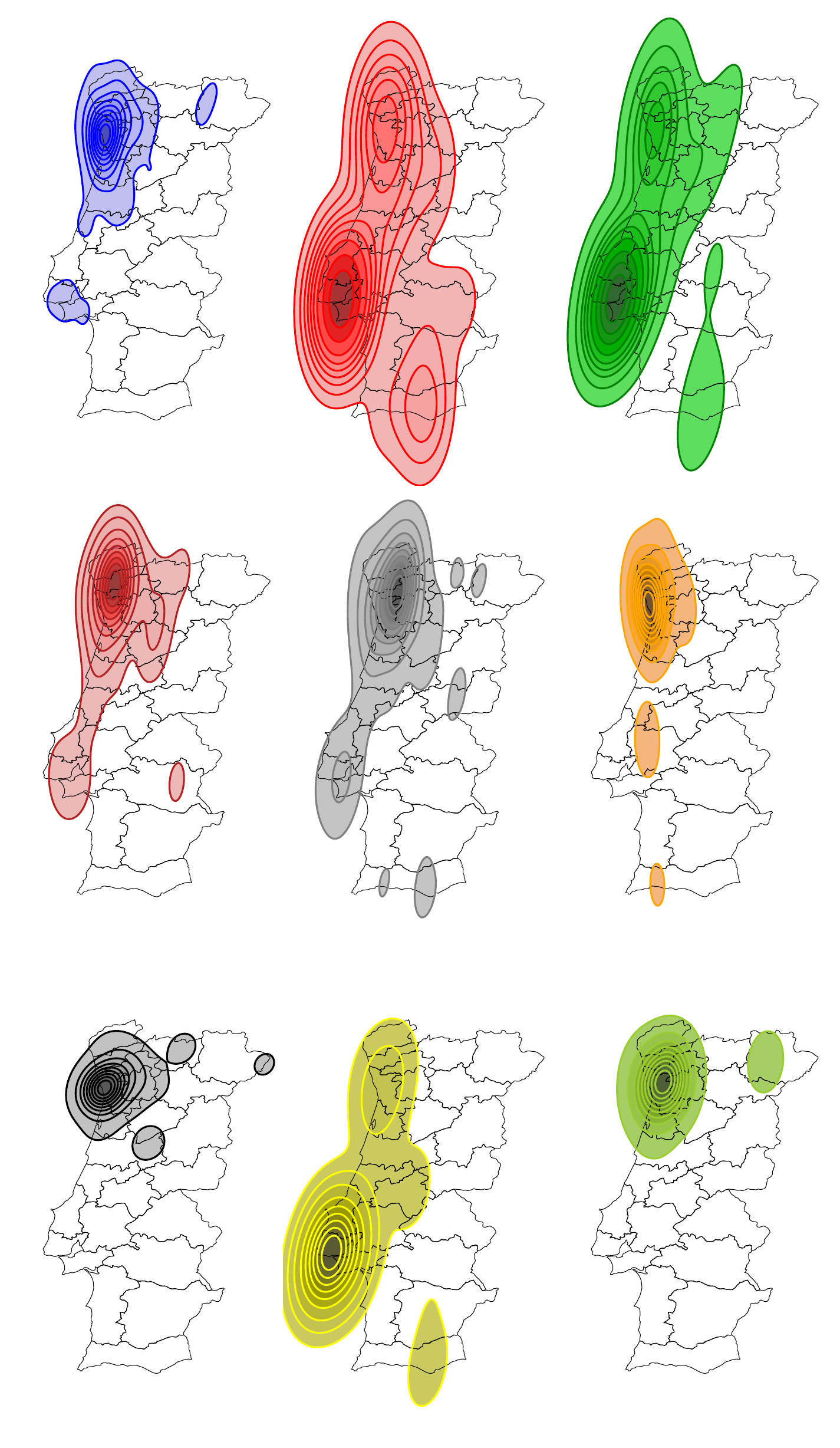}
\caption{The KDE maps of the top 9 club academies in Portugal, measured by the number of youth team players currently at a senior level in the Portuguese pyramid. The clubs are (from right to left, top to bottom) FC Porto, SL Benfica, Sporting CP, Sporting Braga, Vitória SC, Rio Ave FC, Boavista, Estoril Praia and Paços de Ferreira.}
\label{zerozero_formacao}
\end{figure*}

\section{Discussion}

The high-level recruitment strategy of the Portuguese clubs is evident in the maps. SL Benfica and Sporting CP cover the most populous parts of the country. Intermediate clubs (e.g., SC Braga, Vitória SC, and Estoril Praia) can recruit from outside their surrounding region but are much more limited than large clubs. The remaining clubs focus their recruitment exclusively on the regions where they are located, with sporadic recruitments in other places (i.e., being more the exception than the rule).

FC Porto’s case is interesting. Despite being a large club, FC Porto does not recruit across the whole country territory. The geographical coverage of FC Porto youth recruitment is below the level of SC Braga and Vitória SC. The recruitment strategy compared to other clubs at the same dimension level seems quite different. Nonetheless, Porto can still produce a high amount of top-level talent, perhaps indicating that they are better at developing than recruiting talent. As stated before, Porto did not build a youth academy with accommodation facilities, opting for an alternative strategy based on the investment in human capital expertise (e.g., talented Portuguese and foreign coaches). Such limitation makes it harder even for big clubs (i.e., FC Porto) to attract talent from more distant locations. However, many environmental variables linked to player development do not require infrastructure, e.g., alignment of expectations and communication\cite{gangso_talent_2021}, allowing clubs to excel without outsized investments.

Estoril Praia is another curious case. Not as widely recognized as a youth development powerhouse, it manages to enter the clubs with an intermediate level of coverage. The club is clearly above the remaining clubs regarding geographical coverage, having a relatively good recruitment level outside the club’s neighborhood.

The main demographic factor for a player to be a professional footballer in Portugal is apparent: being born in a high-density place correlates greatly with where top-tier soccer clubs are located. In fact, the main driver for talent in a given region is how many clubs are there and how good their academy is. This leads to a vicious cycle where clubs with better financial conditions have the resources to invest in their local academies. Correspondingly, the investment produces outsized returns, increasing the clubs’ dominance.

The maps presented in this paper show that the northern coastal and the Lisbon regions are extensively explored. However, the two areas must be examined by comparing the maps in Figure \ref{zerozero_formacao}. The coastal cities between Porto, Lisbon, and the Algarve region produce fewer footballers than expected due to their population density. There is a big reason for this: most of the major clubs in these regions (Beira-Mar SC, Académica OAF, UD Leiria, …) are currently in lower-tier, non-professional leagues. This leads to a lack of investment in their academies, leaving many potential talents without the proper resources to develop and mature.

Bigger clubs are managing to fill some of these gaps. For example, in Figure \ref{zerozero_formacao}, we see that Benfica reaches the third density level in Algarve. By detecting this underserved hotspot and having training staff and infrastructure in these areas, Benfica is recruiting talent from Algarve with very little competition from local teams.

However, there is no substitute for having professional teams in the region. As we see from the maps of the more local-focused clubs, they obtain a very high density across their surroundings. This is essential for the Portuguese pyramid since it allows talent with slower maturity to maintain a good competitive level to reach the professional level at later stages of their career. 

Smaller clubs in Portugal are unable to compete with the Big Three. Financially, the difference is massive: in contrast to the top 5 leagues where the income ratio in television rights between the most significant and the median earners varies from 1.2x to 2.9x, Portuguese clubs have an income ratio of 9.2x \cite{uefa_european_2022}, by far the worst in UEFA. This financial capacity extends to their youth academies, where top clubs can afford full facilities and coaches and those with less income. Perhaps more important than this, the top clubs can poach the talents of other youth teams, leaving them without the possibility of developing young players themselves, losing most of the investment. In sum, smaller clubs can’t compete financially with the Big Three, which has implications for youth talent retention.

This raises the question: how can smaller clubs cope with bigger clubs’ capacity to attract talent? There are three ways to do this: (1) focusing on local talent, providing a less disruptive path for local players to develop without requiring moving to a big academy with all the associated challenges (i.e., less social support from the most relevant caring figures, having to adapt to a new school, etc.), (2) offering a clear pathway to the first team, appealing to young players that might otherwise have lower odds of succeeding in a big clubs’ academy, and (3) partnering with the big clubs, allowing them to get access to underutilized talent. Despite FC Porto’s success with its more local strategy, other clubs will not be able to replicate their success because they do not have the resources to recruit and maintain top-tier coaches in their youth academies. But similarly to developing young players, young coaches can also be developed in smaller clubs, allowing them to close the gap in human resources quality compared to the top team.

\section{Conclusion}
Youth academies in Portugal are increasing their returns massively over the last decades. Small investments in infrastructure and staff are returning millions in transfer revenue. As one of the significant components of a youth academy, it is imperative to have an adequate youth talent intake strategy for the success of a club.

Data is a massive asset in improving talent intake. Our characterization of how youth players are distributed across the country for each team provides a straightforward way to describe the strategy of football teams regarding youth recruitment.

Although restricted to public datasets, we discovered several components of clubs’ strategies. And more importantly, we identified possible avenues for improving youth talent intake and development that clubs can explore to improve their academy results.



\begin{funding}
This work is partially financed by National Funds through the Portuguese funding agency, FCT - Fundação para a Ciência e a Tecnologia, within project UIDB/50014/2020.
\end{funding}

\section{Endnotes}

\bibliographystyle{SageV}
\bibliography{my.bib}

\end{document}